\documentclass[12pt]{article}
\usepackage[left=2cm, right=2cm, bottom=2cm ,top = 2cm, footskip=1cm]{geometry}
\usepackage[font=small,labelfont=bf]{caption}
\usepackage{multirow}
\usepackage{setspace}
\usepackage{tabularx}
\usepackage{lineno}

\let\counterwithin\relax
\usepackage{chngcntr}
\usepackage{float}
\usepackage{hyperref}
\usepackage{tcolorbox}

\usepackage[authoryear,sort&compress]{natbib}
\usepackage{natbib}
\setcitestyle{open={(},close={)}}

\usepackage{graphicx}
\usepackage{amsmath}

\usepackage[section]{placeins}

\title{Identifying epileptogenic abnormality by decomposing intracranial EEG and MEG power spectra}
\author{Csaba Kozma$^{1*}$, Gabrielle Schroeder$^{1}$, Tom 
 Owen$^{1},$ Jane de Tisi$^{3}$,\\  Andrew W. McEvoy$^3$,  Anna Miserocchi$^{3}, $ John Duncan$^{3}$,\\ Yujiang Wang$^{1,2,3}, $ Peter N. Taylor$^{1,2,3}$}

\begin{document}

\maketitle

\begin{enumerate}
\item{CNNP Lab (www.cnnp-lab.com), Interdisciplinary Computing and Complex BioSystems Group, School of Computing, Newcastle University, Newcastle upon Tyne, United Kingdom}
\item{Faculty of Medical Sciences, Newcastle University, Newcastle upon Tyne, United Kingdom}
\item{UCL Queen Square Institute of Neurology, Queen Square, London, United Kingdom}
\end{enumerate}

\begin{center}
* c.a.kozma2@newcastle.ac.uk    
\end{center}

\footnotesize
\begin{singlespace}
HIGHLIGHTS:
\begin{itemize}
  
  \item  Interictal EEG biomarkers using normative maps of power in different frequencies allow us to identify epileptic abnormalities and distinguish patient outcome.
  \item Decomposing brain activity into periodic (oscillatory) and aperiodic (trend across all frequencies) activity may identify underlying drivers of epileptic abnormalities.
  \item Our findings suggest that sparing abnormalities in either periodic or aperiodic activity may lead to a poor surgical outcome, henceforth both are necessary.
\end{itemize}

\end{singlespace}

\newpage

\section{Abstract}
    \fontsize{10pt}{12pt}\selectfont

\textbf{Background}

Identifying abnormal electroencephalographic activity is crucial in diagnosis and treatment of epilepsy. Recent studies showed that decomposing brain activity into periodic (oscillatory) and aperiodic (trend across all frequencies) components may illuminate drivers of changes in spectral activity. 

\textbf{Methods}

Using iEEG data from 234 subjects, we constructed a normative map and compared this with a separate cohort of 63 patients with refractory focal epilepsy being considered for neurosurgery. The normative map was computed using three approaches: (i) relative complete band power, (ii) relative band power with the aperiodic component removed (iii) the aperiodic exponent. Corresponding abnormalities were also calculated for each approach in the separate patient cohort. We investigated the spatial profiles of the three approaches, assessed their localizing ability, and replicated our findings in a separate modality using MEG.

\textbf{Results}

The normative maps of relative complete band power and relative periodic band power had similar spatial profiles. In the aperiodic normative map, exponent values were highest in the temporal lobe. Abnormality estimated through the complete band power robustly distinguished between good and bad  outcome patients. 
Neither periodic band power nor aperiodic exponent abnormalities distinguished seizure outcome groups. Combining periodic and aperiodic abnormalities improved performance, similar to the complete band power approach (iEEG AUC=0.64, p=0.05; MEG AUC=0.69, p=0.039).

\textbf{Conclusions}

Our findings suggest that sparing cerebral tissue that generates abnormalities in either periodic or aperiodic activity may lead to a poor surgical outcome. Both periodic and aperiodic abnormalities are necessary to distinguish patient outcomes, with neither sufficient in isolation. Future studies could investigate whether periodic or aperiodic abnormalities are affected by the cerebral location or pathology.

\fontsize{12pt}{10pt}\selectfont
\newpage

\section{Introduction}
\doublespacing
\setlength\parindent{24pt}

Improved EEG biomarkers of the epileptogenic zone are important, as around half of individuals have recurrent seizures after surgical treatment \citep{Téllez2005,deTisi2011,Wiebe2001}. Substantial research has focused on interictal EEG biomarkers using normative maps of power in different frequencies \citep{Bernabei2022,Frauscher2018,Taylor2022}. This approach involves outlining the spatial characteristics and ranges of the feature of interest in a healthy context and comparing patient data to identify abnormalities. The invasive nature of intracranial EEG (iEEG) means that obtaining data from healthy individuals for comparison is not possible. Recent studies proposed using patient iEEG data from regions outside the epileptogenic zone to create normative maps \citep{Betzel2019,Frauscher2018,Groppe2013}. The band power of different frequency bands can be used to infer expected healthy spatial profiles of EEG activity. This approach showed promising results to identify abnormalities and classify patient outcomes across different modalities \citep{Bernabei2022,Owen2023,Taylor2022}.

It is unclear what spectral features drive the observed band power abnormalities. One way to explore these features is to decompose the power spectra into rhythmic (periodic) and non-rhythmic (aperiodic) components \citep{Buzsaki2012,Gerster2022,He2014,Miller2009}. The aperiodic component can be described by its offset and exponent \citep{Buzsaki2012, Miller2009}, while periodic components form peaks in the power spectra. Both components should be considered, as power changes observed in specific frequency bands can be attributed to either changes in the peak or the offset \citep{Donoghue2022,Donoghue2020b, Gao2017,Gerster2022}.

Recent empirical studies have related periodic and aperiodic components to behavioral and demographic variables. There is an association between aperiodic activity and age \citep{Donoghue2020b,Voytek2015,Voytek2015b}. Aperiodic activity varies across neural-development and decline \citep{Ostlund2022,Tran2020} as well as in sleep \citep{Lendner2020} and anesthesia \citep{Colombo2019}. Despite the growing interest in decomposing power spectra \citep{Donoghue2020,Gerster2022, Ostlund2022,Wilson2022}, previous studies have not investigated which power spectrum components contribute to abnormal interictal brain activity.

Here we investigate if interictal abnormality is driven by periodic components, aperiodic components, or a combination of both components. We first present normative maps of complete band power, periodic band power, and the aperiodic exponent obtained from interictal intracranial EEG (iEEG) recordings from 234 participants. Subsequently, we identify abnormalities in spared and resected regions using recordings obtained from 63 patients with focal epilepsy. Lastly, we replicate our analysis using resting-state magnetoencephalography (MEG) recordings from 70 healthy controls and 33 patients. Ultimately, we aim to clarify if periodic and/or aperiodic abnormalities explain overall band power abnormality, and if they are both necessary to distinguish patient outcomes.

\section{Methods}

\subsection{Patients \& controls}
\subsubsection{iEEG cohorts}

We generated normative maps using data from 234 controls from the Restoring Active Memory (RAM) data set \citep{RAM2018}. The second cohort was from 63 patients with refractory focal epilepsy treated at National Hospital for Neurology and Neurosurgery (NHNN) (Table 1). Surgical outcomes of seizure freedom for the UCLH cohort were assessed using the International League Against Epilepsy (ILAE) surgical outcome classification, focusing on the 12-month post-surgery period. The NHNN cohort consisted of 34 $ILAE_{1,2}$ and 29 $ILAE_{3+}$  patients, as described previously \citep{Taylor2022}.

\captionof{table} {Summary of iEEG cohort data} \label{tab:title}
\begin{center}
\begin{tabular}{ |p{8cm}|p{2.1cm}|p{2.1cm}|p{5cm}|} 

\hline
 & $ILAE_{1,2}$  & $ILAE_{3+}$ & Test statistic\\
\hline
\multirow{5}{22em}{n\% \\ Age, mean (SD) \\ Sex, male:female \\ Temporal, extratemporal \\ Resection hemisphere, left/right \\ Number of electrode contacts, mean (SD)} &  34 (54)  & 29 (46) &\\ 
& 32.3 (10.7) & 33 (8.8) &  p = 0.801, t = -0.252\\ 
& 16:18 & 17:12  & p=0.361, $X_{2}$=0.838\\
& 22, 12 & 15, 14  & p=0.296, $X_{2}$ = 1.088\\
& 19/15 & 16/13  & p= 0.955, $X_{2}$ = 0.003\\
& 67.5 (29.27) & 65.9 (23.3)  & p= 0.459, t= -0.103\\
\hline
\end{tabular}
\end{center}

\subsubsection{MEG cohorts}

We conducted the same analysis on magnetoencephalography (MEG) recordings using healthy controls and patients. Both groups underwent eyes-closed awake resting-state MEG using a 275-channel CTF whole head MEG system in a magnetically shielded room. Healthy control data were collected at CUBRIC, Cardiff, consists of 70 participants. Data from 33 patients with refractory neocortical epilepsy were collected at NHNN (Table 2). This patient cohort consisted of 12 $ILAE_{1,2}$ and 21 $ILAE_{3+}$ patients, as described previously \citep{Owen2023}.

\captionof{table} {Summary of MEG cohort data} \label{tab:title}
\begin{center}
\begin{tabular}{ |p{8cm}|p{2.1cm}|p{2.1cm}|p{5cm}|} 
\hline
 &  $ILAE_{1}$  & $ILAE_{2+}$ & Test statistic\\
\hline
\multirow{4}{22em}{n\% \\ Age, mean (SD) \\ Sex, male:female \\ Resection hemisphere, left/right } &  12 (36)  & 21 (64) &\\ 
& 32.3 (10.7) & 32.3 (11.3) &  p = 0.058, t = 2.003\\ 
& 7:5 & 10:11  & p=0.554, $X_{2}$=0.351\\
& 6/6 & 12/9  & p= 0.692, $X_{2}$ = 0.157\\
\hline
\end{tabular}
\end{center}

\subsection{MRI processing}
Pre-operative 3D T1-weighted MRI images were acquired using a 3T GE Signa HDx scanner using previously described acquisition sequences \citep{Taylor2018}. These MRI data were used to segment and parcellate different brain regions using the FreeSurfer ‘recon-all’ pipeline \citep{Fischl2012}. For consistency between subsequent iEEG and MEG analyses, we used a common Lausanne parcellation with 114 neocortical brain regions \citep{Hagmann2008}. For iEEG analyses we also included 14 deep brain regions including hippocampus, amygdala, thalamus, putamen and caudate. All time series analysis was performed at a brain region level, rather than at a iEEG electrode contact or MEG sensor level.\\

For iEEG data, we localized electrode contacts to the nearest brain regions, based on the Euclidean distance. Electrode contacts further than 5mm from a region, or located within white matter and more than 2mm from grey matter were excluded, as previously \citep{Taylor2022}.\\

For MEG data, we applied the minimum norm estimate technique, sLORETA, in conjunction with an overlapping spheres head model, to perform source localization, projecting onto the cortical surface \citep{Tenney2020}. This process yielded 15,000 sources that were constrained in a direction perpendicular to the cortex. Sources were downsampled into scout regions of interest, using Lausanne parcellation.\\

To generate resection masks we registered the postoperative T1-weighted scan to the preoperative scan using linear registration. The resected tissue was manually delineated with FSL software \citep{Jenkinson2012}, accounting for anatomical information (e.g. due to post-operative brain shift into the resection cavity). Electrode contacts within a 5mm proximity to the resection were labeled as resected. Regions were classified as resected if the volumetric change between the pre- and postsurgical measurements exceeded 25\%.

\subsection{iEEG processing}

We used the same data as \citep{Taylor2022} and followed those preprocessing steps. The RAM and UCLH data were downsampled to 200 Hz and a common average reference was then applied to all recordings. For the RAM cohort, channels in seizure onset zones (SOZ), early propagation zones, brain lesions, or white matter were removed, resulting in 21,598 channels across 234 participants. For the UCLH cohort, we retrospectively extracted 70 seconds of interictal iEEG recordings at least 2 hours away from any seizures. Grey matter channels without recording artifacts were included, resulting in 4,273 channels across 63 patients.

\subsection{MEG processing}

We used eyes-closed, awake resting state MEG preprocessed data \citep{Owen2023}. In brief, raw MEG recordings were preprocessed using Brainstorm \citep{Tadel2011}. MEG sensor locations and structural MRI scans were aligned and registered using fiducial points. The recordings were down sampled to 600 Hz and bandpass filtered from 1 to 100 Hz. Cardiac and ocular artifacts were removed using independent component analysis (ICA). Subsequently, the sources were downsampled into neocortical ROIs, using the Lausanne parcellation. We excluded deep brain subcortical regions as it is challenging to determine the neural activity or sources within the brain that give rise to the measured magnetic fields outside the head, especially for interictal recordings expected to be of low amplitude \citep{Ahlfors2010,Baillet2001}.

\subsection{Estimating power spectra}
Based on the preprocessed iEEG and MEG time-series data we estimated the complete power spectrum in each 70 second epoch. Using Welch’s method, we divided the time-series data into 2-second temporal windows with 1-second overlap, applied a Hamming window to each temporal window, and computed the Fourier transform, and then averaged the resulting spectra.

\subsection{Decomposing power spectra}

To decompose the iEEG and MEG power spectra, we used the specparam package \citep{Donoghue2020b}. The algorithm was used with configurations of ‘peak width limits’ = [1,8], ‘maximum number peaks’ = 6, ‘min peak height’ = 0.1, ‘peak threshold’ = 0.1, ‘aperiodic mode’ = ‘fixed’, and ‘frequency range’= [1,30] Hz. We only used the 1-30 Hz frequency range to exclude issues with line noise and possible errors in curve fitting in higher frequencies. The decomposition sequentially fit aperiodic and periodic components to the neural power spectrum. First, specparam performed a robust linear fit of the spectrum to estimate the aperiodic signal (Fig. 1A) \citep{Donoghue2022,Donoghue2020b, Gerster2022}. The fitted curve was then subtracted from the original spectrum, resulted in a flattened spectrum representing periodic activity (Fig. 1A). A relative threshold was calculated based on the standard deviation of the flattened spectrum. Peaks exceeding predefined thresholds were identified iteratively, and Gaussian functions were fitted to these peaks. The fitted Gaussian curves were subtracted from the spectrum. The oscillatory components were obtained by fitting a multivariate Gaussian to all the extracted peaks simultaneously. Finally, the initial fit is added back to the flattened spectrum, resulting in the periodic component. In addition to analyzing the entire power spectrum, we focused on two specific decomposed features: periodic activity and aperiodic exponent.
We calculated the complete band power and periodic band power within four frequency bands of interest ($\delta$ 1–4 Hz, $\theta$ 4–8 Hz, $\alpha$ 8–13 Hz, and $\beta$ 13–30 Hz). The complete and periodic band power values were then $log_10$ transformed and normalized, ensuring that the sum of the band power in each contact (for iEEG data) or ROI (MEG data) equaled one (Fig. 1, side views of the maps in Fig S1.1). In addition, we extracted aperiodic exponent values. These steps yielded the relative complete band power, relative periodic band power, and aperiodic exponent values for each subject in each of the subject’s contacts (iEEG) or ROIs (MEG).

\begin{figure}[H]
	\centering
	\includegraphics{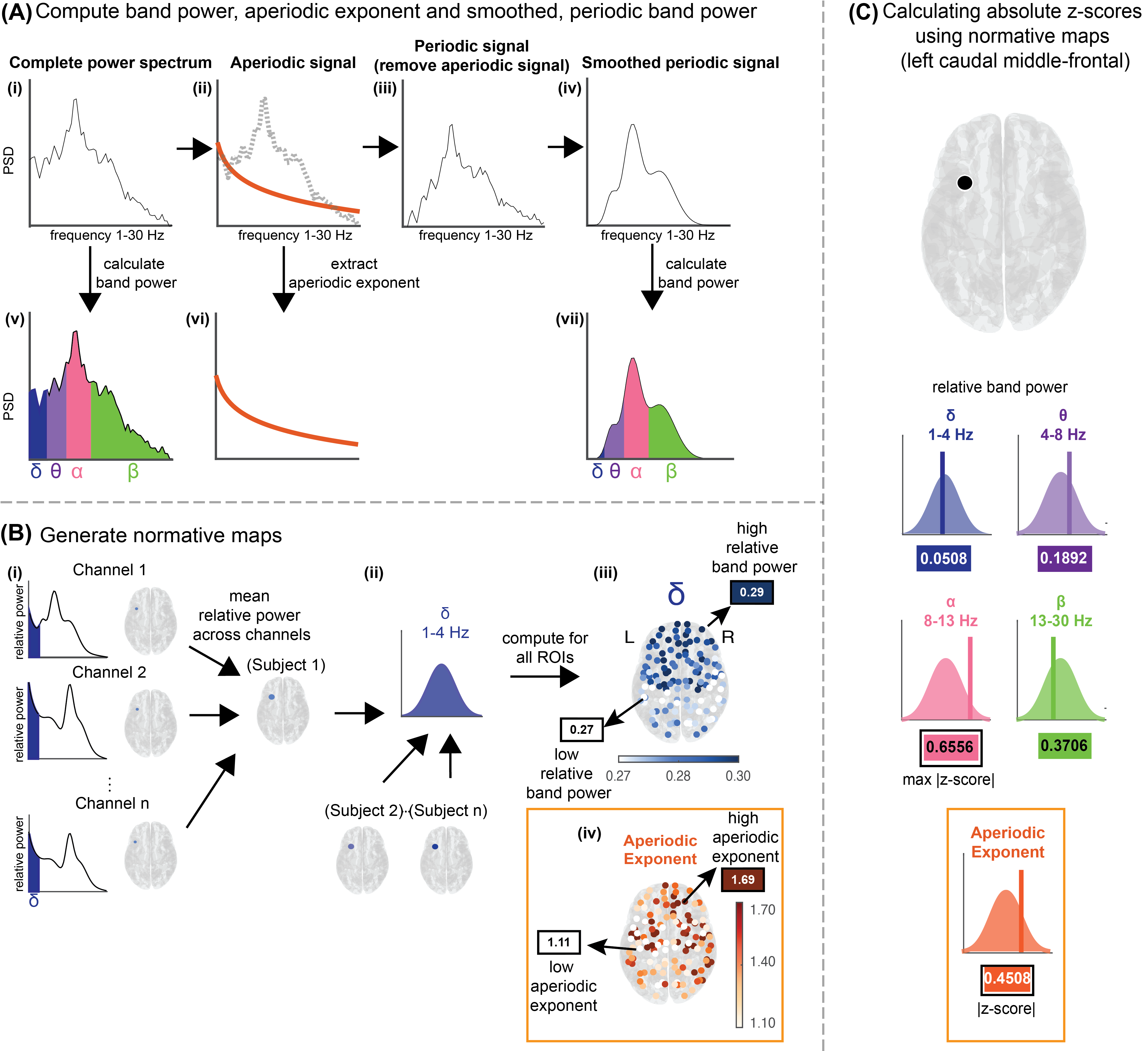}

\captionof{figure}{\textbf{Computing abnormalities based on the decomposed power spectrum.} (A) Decomposition of iEEG and MEG power spectra using specparam to compute relative complete band power, the aperiodic exponent, and relative periodic band power. (i) Complete power spectrum between 1-30 Hz. The power spectrum (ii, dotted grey line) was decomposed in to a periodic component (iii, orange line) and periodic signal (iv, Smoothed periodic signal). (v) Complete band power was calculated for four frequency bands ($\delta$ 1–4 Hz, $\theta$ 4–8 Hz, $\alpha$ 8–13 Hz, and $\beta$ 13–30 Hz). (vi) Aperiodic exponent values were extracted (orange line). (vii) Periodic band power was calculated for four frequency bands. 
(B) Using $\delta$ band as example to show how normative maps are generated. (i) Mean relative complete band power/relative periodic  band power/aperiodic exponent were computed across contacts to get a single series of values per  ROI ($\delta$ band as example). We then computed the measures for every ROI (ii) across normative map subjects. (iii) We generated the normative map, which is the spatial distribution of relative complete band power/relative periodic  band power across the whole brain. (iv) We generated aperiodic exponent maps. 
(C) Absolute z-scores calculated across the frequency bands and the maximum was taken for every ROI, which serves as the value of abnormality.
} 
\label{Methods}
\end{figure}

\subsection{Generate normative maps}

iEEG normative maps of relative log band power values were generated for the complete power spectrum and the periodic activity. We averaged this measure across electrodes within subjects first, then across subjects within each region to compute the mean ($\mu$) and standard deviation ($\sigma$) across subjects (eq. 1). For the aperiodic exponent values, we calculated the mean and standard deviation of exponent values across electrodes and patients within each region (Fig. 1B and Fig. 2).
MEG normative maps were generated via the same method as described above and can be found in the supplementary material (Fig. S1.2).

\subsection{Calculate abnormalities using normative maps}

To assess the abnormality of a region’s relative band power in a patient, relative to the normative map, we calculated the absolute z-score of each region $i$ in each frequency band $j$ (see also Fig. 1):

\begin{equation}
|z_{ij}| =  \left|\frac{x_{ij}-\mu_{ij}}{\sigma_{ij}}\right|        
\end{equation}

\vspace{10mm}

Here $x$ represents the regional  band power value for an individual patient and $\mu$ and $\sigma$ the mean and standard deviations of the regional band power in the normative map.  Meanwhile, $i$ represents the region, and $j$  the frequency band of interest. We used the maximum absolute z-score across frequency bands to define abnormality at the regional level, considering individual differences in frequency bands both in complete band power and periodic band power (Fig. 1C). Regarding aperiodic exponent, we calculated the absolute z-score of each region in the patient data set relative to the normative map. 

To compare the values between resected and spared regions, we used the distinguishability statistic ($D_{RS}$) for each patient in the data set and frequency band \citep{Bernabei2022,Ramaraju2020,Wang2020}. The $D_{RS}$ value indicates the abnormality difference between resected and spared regions: a value above 0.5 indicates that spared regions had higher abnormality than resected regions, while a value below 0.5 indicates the opposite. $D_{RS}$ is equivalent to the area under receiver operating curve. 

\subsection{Ethics approval}

The analysis of this data set was approved by the Newcastle University Ethics Committee (ref. 12721/2018).

\subsection{Code and data availability}

Code to reproduce figures in the manuscript will be made available upon acceptance of the paper. 
Deidentified PSD data will also be made available upon acceptance of the paper.

\section{Results}

We decomposed the power spectrum into its periodic and aperiodic components to investigate these components as potential driving mechanisms of band power abnormalities in the complete power spectrum. First, we show the normative maps of (i) complete band power, (ii) periodic band power, and (iii) the aperiodic exponent using interictal intracranial EEG (iEEG) recordings from 234 subjects. Second, we compared the abnormalities of spared and resected regions in two example patients across the three approaches. We used the $D_{RS}$ metric to quantify the relationship between spared and resected abnormalities. Third, we compared the $D_{RS}$ values computed from the abnormalities calculated from complete band power, periodic band power, and aperiodic exponent in interictal iEEG recordings acquired from 63 patients with focal epilepsy. Finally, we repeated our analysis using resting-state MEG recordings.

\subsection{Normative maps of complete band power, periodic band power, and aperiodic exponent}
We first show the spatial distribution of healthy neural activity based on complete and the decomposed spectral features. Figure 2A presents normative maps of the relative complete band power in four distinct frequency bands ($\delta$ 1–4 Hz, $\theta$ 4–8 Hz, $\alpha$ 8–13 Hz, and $\beta$ 13–30 Hz). We then generated normative maps based on relative periodic band power, with the aperiodic component removed (Fig. 2B). These maps reveal several discernible patterns. Notably, the anterior temporal and frontal regions have the highest relative power in the delta frequency band, whereas the parietal and occipital regions show prominent relative power in the alpha frequency band. Thus, periodic band power reproduces the previously observed spatial patterns in complete band power. Figure 2C shows that the normative maps of aperiodic exponent values also have a specific spatial pattern; namely, higher aperiodic exponent values can be found in the temporal lobe compared to the rest of the brain. These results are replicated using MEG data in Supplementary Section/Figure S2.

\begin{figure}[H]
	\centering
	\includegraphics{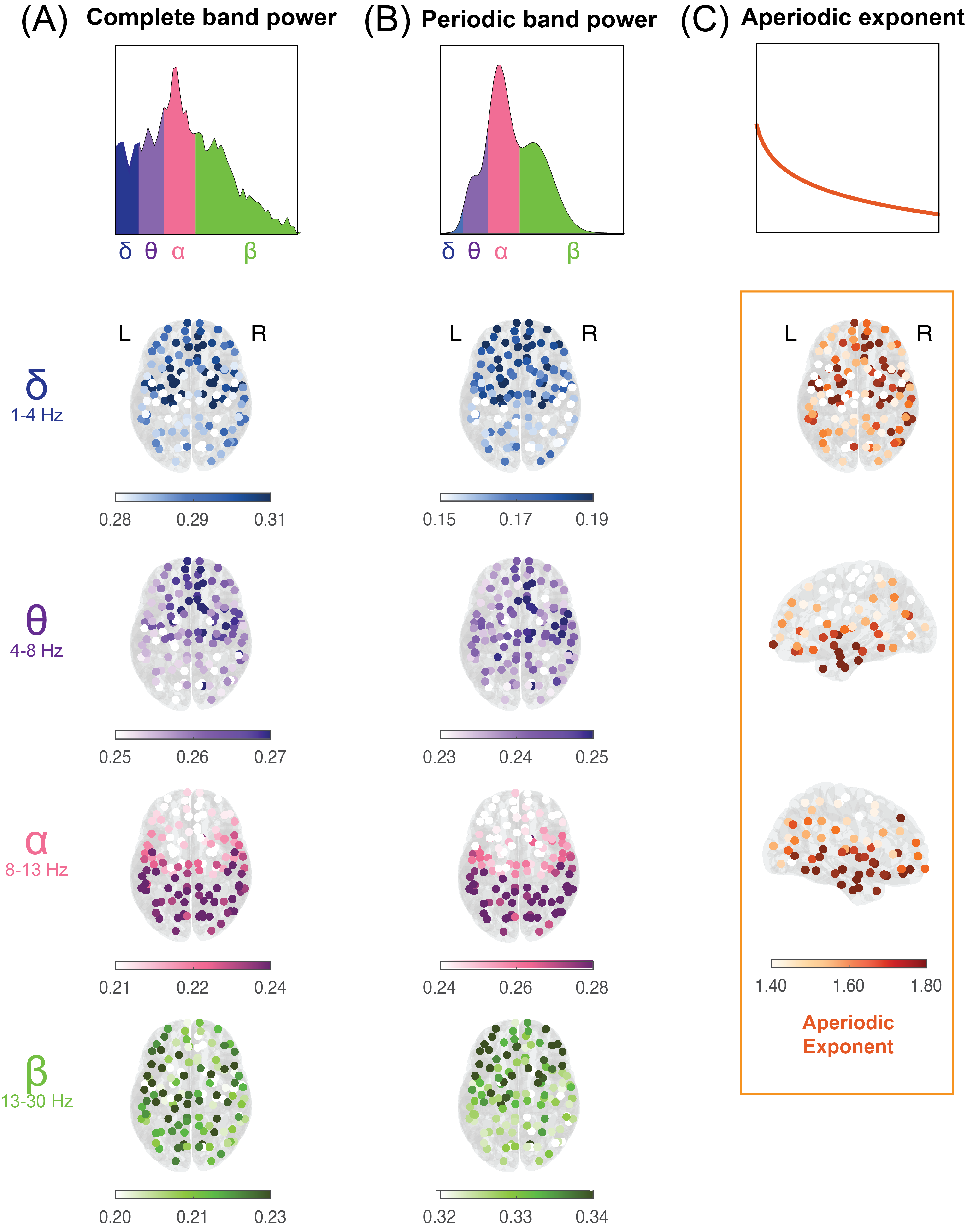}
	\caption{\textbf{Normative maps of iEEG relative complete band power, relative periodic  band power, and aperiodic component.} Spatial distribution of mean ROI values of, (A) relative complete band power, (B) relative periodic  band power, and (C) aperiodic exponent. The colour axes scale differs for each frequency band and the distribution of exponent values. Top row shows the schematic representation of each of the approaches. 
	}
	\label{Normative maps}
\end{figure}

\subsection{Different complete band power components contribute to abnormality in different patients }

We computed abnormalities in complete band power, periodic band power, and the aperiodic exponent in two example patients (Fig. 3). The abnormalities capture the largest deviation of the spectral measure from the normative distribution in each region of interest (ROI). In addition, we computed $D_{RS}$, the distinguishability of resected vs. spared tissue, from each type of abnormality.
Figure 3A shows the abnormalities of an example patient who was seizure free after surgical intervention (ILAE 1). Both complete band power and periodic band power abnormalities yielded $D_{RS}$ values well below 0.5 ($D_{RS}$=0.14, $D_{RS}$=0.26) and were strongly correlated in this patient (Spearman’s correlation=0.90, Figure 3C). However, the aperiodic exponent abnormalities gave $D_{RS}$ value close to 0.5 ($D_{RS}$=0.44), and were not informative for outcome nor correlated to our original measure (Spearman’s correlation=0.29). Thus, the complete band power gained from the periodic signal performed better than the aperiodic exponent.\\
In contrast Patient 2 had poor outcome after surgery (ILAE 4)(Figure 3B). Both complete band power and aperiodic component abnormalities yielded $D_{RS}$ values above 0.5 ($D_{RS}$=0.94, $D_{RS}$=0.84), and these abnormalities were strongly correlated (Spearman’s correlation=0.72, Figure 3D). However, the periodic band power abnormalities gave $D_{RS}$ value close to 0.5 ($D_{RS}$=0.55) and indicating similar abnormalities in spared and resected ROIs (Spearman’s correlation=0.23). The aperiodic exponent solution best estimated abnormalities and patient outcome for Patient 2. Therefore, the type of component that appeared more informative about the location of abnormality and outcome of patient differed between these two patients.

\begin{figure}[H]
	\centering
	\includegraphics[width=\textwidth]{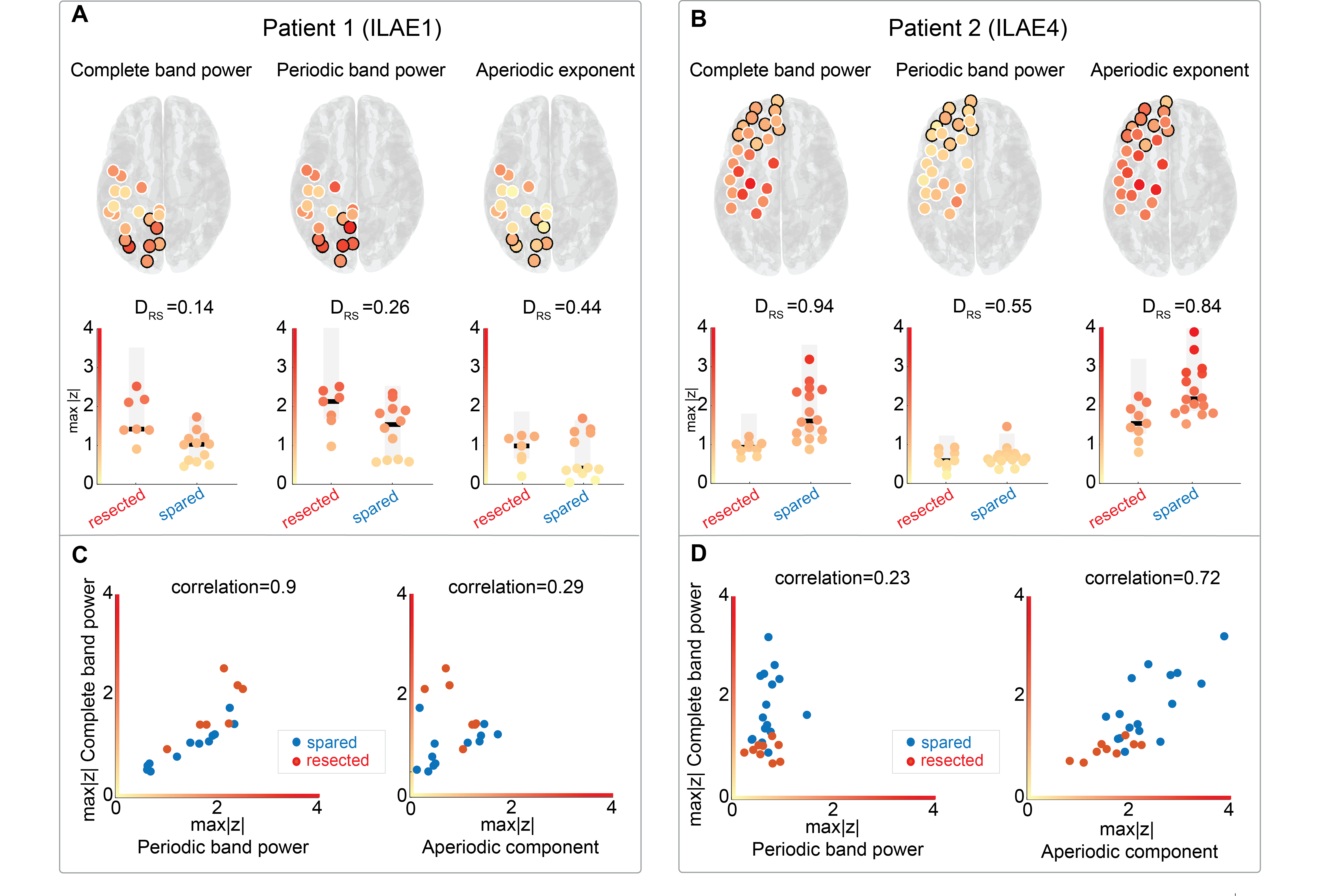}
	\caption{\textbf{Different contributors to abnormality identified across} relative complete band power, relative periodic band power and aperiodic exponent as a marker of resected and spared tissue in two example individual patients. Postoperative regions that were later surgically resected are circled in black. Non-resected regions are circled in white. (A) Patient 1 did not show signs of continued postoperative seizures (ILAE1), (B) Patient 2 did (ILAE4). The type of component that is most informative about the location of abnormality and outcome of patient appears to vary across patients. (C) ILAE1 patient, complete band power and periodic band power abnormalities correlate strongly, (D) ILAE4 patient show strong correlation between complete band power and aperiodic exponent abnormalities. }
	\label{Normative maps}
\end{figure}

\subsection{iEEG: Periodic and aperiodic features identify different parts of abnormality}

The previous section suggests different factors can drive abnormalities in two patients, subsequently affecting $D_{RS}$. We next explored these factors across the larger cohort.

In the whole cohort, Figure 4A shows that abnormality estimated through the complete band power clearly separates good and bad outcome patients (AUC=0.71, p$<$0.01). However, neither periodic band power abnormalities (Fig. 4B) (AUC=0.57, p=0.23) nor aperiodic exponent abnormalities (Fig. 4C) (AUC=0.56, p=0.27) alone separated patient groups.

Figures 3 and 4ABC indicate that abnormalities may be driven by different components in different patients, in different regions. We therefore next took an agnostic approach by computing the maximum abnormality, irrespective of the component used (periodic or aperiodic). In this approach we next selected the largest of the periodic and aperiodic abnormalities for each ROI (Fig. 4D). The $D_{RS}$ values of these maximum abnormalities improved outcome classifications (AUC=0.64, p=0.05).

Complete band power $D_{RS}$ values were only weakly correlated to the periodic band power $D_{RS}$ values (Fig. S2.1A, Spearman’s correlation=0.54) and aperiodic exponent $D_{RS}$ values (Fig. S2.1B, Spearman’s correlation=0.37). However, $D_{RS}$ values computed from the maximum abnormalities were more strongly associated with the complete band power $D_{RS}$ values (Fig. S2.1C, Spearman’s correlation=0.71).

Thus, both periodic and aperiodic abnormalities were necessary to more accurately distinguish patient outcomes, with neither sufficient in isolation.

\begin{figure}[H]
	\centering
	\includegraphics[width=\textwidth]{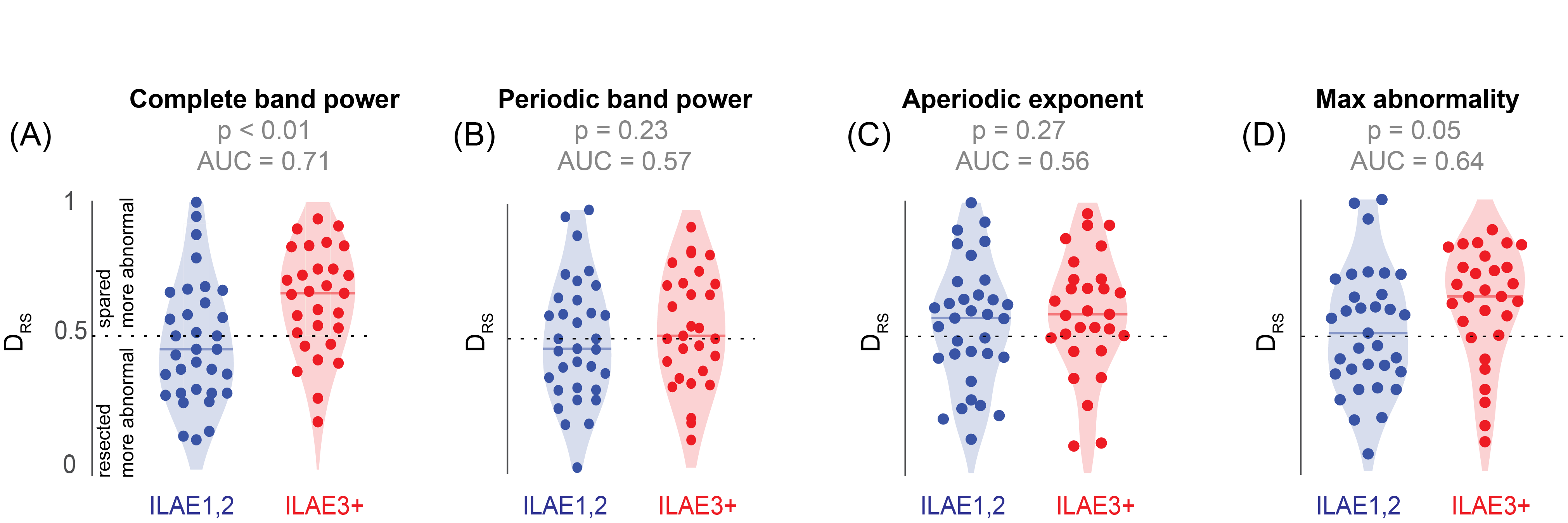}
	\caption{\textbf{Both periodic and aperiodic abnormalities are necessary to distinguish patient outcomes, with neither sufficient in isolation (iEEG cohort).} (A-D) Comparison of $D_{RS}$ values between ILAE$_{1,2}$ and ILAE$_{3+}$ patients using (A) the complete band power abnormalities, (B) periodic band power abnormalities, (C) aperiodic exponent abnormalities, and (D) maximum abnormalities selected from periodic band power and aperiodic exponent abnormalities for each ROI. Each data point in the plot represents an individual patient, while the darker horizontal line denotes the median $D_{RS}$. AUCs were computed from the receiver operator characteristic curve (ROC) using $D_{RS}$ as a binary classifier for surgical outcome. }
	\label{Normative maps}
\end{figure}

\subsection{MEG: Periodic and aperiodic features identify to different parts of abnormality}

We repeated our analysis using resting-state MEG recordings by generating normative maps from 70 healthy controls and computing ROI abnormalities in 33 patients with refractory neocortical epilepsy. As in the iEEG analysis, neither periodic band power (Fig. 5B) (AUC=0.51, p=0.37) nor aperiodic exponent abnormalities (Fig. 5C) (AUC=0.59, p=0.14) distinguished patient outcomes well. However, the other two approaches (Fig. 5A,D) performed substantially better (complete band power AUC=0.69, p=0.03; max abnormality AUC=0.69, p=0.04). Likewise, the complete band power and the periodic band power approaches was moderately correlated (Fig. S2.1D) (Spearman’s correlation=0.52), whereas the aperiodic approach only weakly correlated (Fig. S2.1E) (Spearman’s correlation=0.26). In contrast, the max abnormality approach exhibited a relatively strong correlation (Fig. S2.1F) (Spearman’s correlation=0.62). 

These findings further suggest that neither periodic nor aperiodic abnormalities alone distinguish patient outcomes, but in combination they do. This agrees with the findings from iEEG data. 

\begin{figure}[H]
	\centering
	\includegraphics[width=\textwidth]{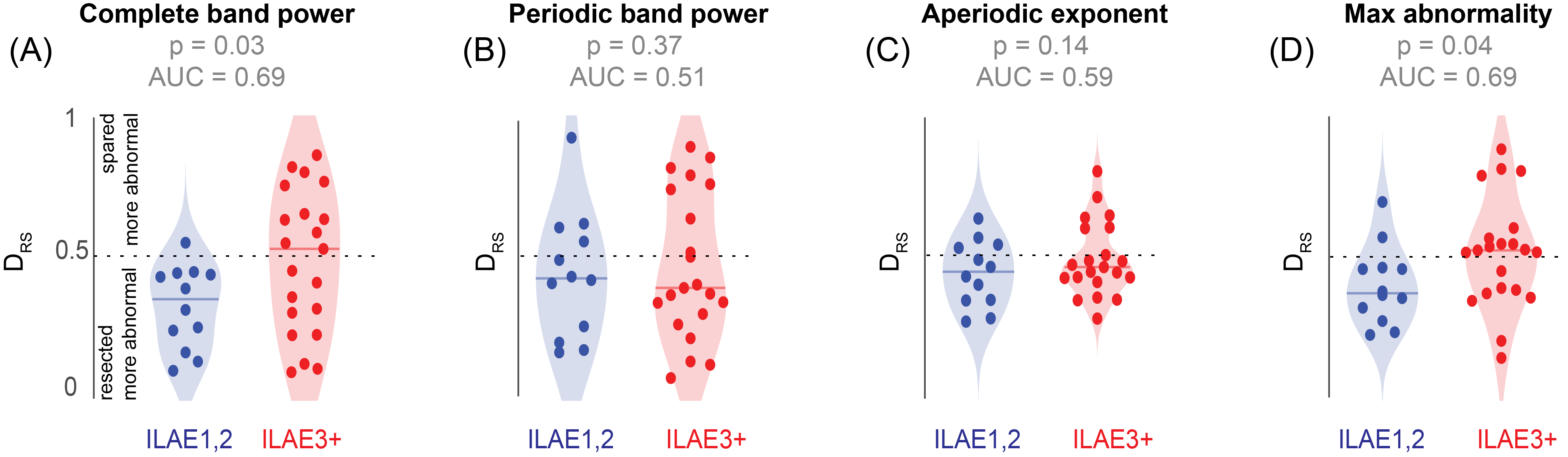}
	\caption{\textbf{Both periodic and aperiodic abnormalities are necessary to distinguish patient outcomes, with neither sufficient in isolation (MEG cohort).} (A) the complete band power abnormalities, (B) periodic band power abnormalities, (C)	aperiodic exponent abnormalities, (D)	maximum abnormalities selected from periodic band power and aperiodic exponent abnormalities for each ROI.}
	\label{Normative maps}
\end{figure}

\section{Discussion}
\setlength\parindent{24pt}
We decomposed iEEG power spectra into periodic and aperiodic components to investigate the driving factors behind previously observed band power abnormalities. Using recordings from 234 subjects, we generated normative maps of complete band power, periodic band power, and the aperiodic exponent. We found that iEEG power spectra abnormalities of different patients were driven by distinct spectral features. Complete band power effectively distinguished between patients with good and bad outcomes, as previously reported \citep{Bernabei2022,Owen2023,Taylor2022}. However, neither periodic band power nor aperiodic exponent abnormalities alone provided the same discriminatory power, whilst their combination does. Repeating the same analysis on resting-state MEG recordings yielded consistent results with the iEEG cohort. Thus, both periodic and aperiodic abnormalities are important for abnormality mapping to predict patient outcomes.

While normative mapping approaches show promise to identify, classify and predict patient abnormalities and outcomes across different modalities \citep{Bernabei2022,Owen2023,Rutherford2023,Sinha2017,Taylor2022,Wang2020}, it is crucial to untangle the drivers of these abnormalities, which may help improve predictions and lead to improved mechanistic understanding of the underlying abnormality. Previous normative mapping approaches investigated how band power abnormality can be used to identify epileptogenic areas and differentiate patient outcome based on resection \citep{Bernabei2022,Owen2023,Taylor2022}. Even if abnormality is identified through this method, the variety of underlying factors must be explored to gain a detailed picture about the nature of epileptogenic abnormality. Decomposing the power spectrum \citep{Gerster2022,He2014,Miller2009,Muthukumaraswamy2018} provides a way to reveal the driving forces and determines the specific spectral components contributing to epileptogenic abnormalities. Toward this goal, we started with decomposing the power spectrum into periodic and aperiodic activity.

The source of aperiodic activity, following a 1/f power law, is highly debated \citep{Buzsaki2012,Donoghue2022,He2014,Miller2009,Miller2014}. The mechanisms behind this phenomenon may involve the combination of the nonlinear dynamics of neural networks, phase coherence-distance relationship between low and high frequency signals, low-pass filtering caused by the dendritic morphology of pyramidal cells, and the captive nature of the extracellular medium \citep{Buzsaki2012,Miller2009,Miller2014}. Other theories attribute the underlying mechanism to a near self-organised criticality \citep{Banerjee2006,Miller2014,Stephani2020}. In addition, \cite{Miller2009} proposed that the power spectral shape is due to the convolution of two functions, both exhibiting exponential decay. These functions represent the post-synaptic current and the membrane leak. Regardless of the exact source of the aperiodic activity, we hypothesized that besides that periodic activity, the contribution provided by the aperiodic component of the power spectrum is substantial.

We generated normative maps for iEEG aperiodic exponent values and found increased values, mainly in the temporal lobe. Although changes to the spatial patterns of aperiodic exponents during task or other activities have been shown in intracranial EEG \citep{Podvalny2015,Sheehan2018}, resting state normative maps have rarely been presented. Exploration of the spatial profile of this normative map requires further investigation. Although the focus of our study was on iEEG data, we also generated normative maps for MEG aperiodic exponent. These maps only partly follow the patterns that have been described previously \citep{Donoghue2020b,Wilson2022}. The differences, namely lower values in the occipital regions, may be caused by the fact that our MEG recordings were eyes closed while previous maps were based on eye open data sets.
The periodic power distribution across the four frequency bands had consistent spatial patterns, mirroring those of complete band power. Similar patterns were found in previous studies \citep{Frauscher2018,Bernabei2022,Groppe2013,Owen2023,Taylor2022}. Namely, the anterior temporal and anterior frontal regions show the highest relative power in the delta frequency band, whereas the parietal and occipital regions demonstrate prominent relative power in the alpha frequency band. We found similar spatial patterns in MEG normative maps for both complete and periodic band power, concordant with previous studies \citep{Donoghue2020b,Owen2023,Wilson2022}.

We found that surgical sparing of cerebral tissue generating abnormalities in either periodic or aperiodic activity was associated with poor surgical outcomes, emphasizing the importance to analyze both types of abnormalities. Comparing abnormalities from periodic and aperiodic approaches and selecting the larger values across all ROIs, improved outcome estimation. This suggests that the driving spectral component of abnormalities varies across regions and patients (see Fig. S3.1, S3.2, S3.3). Future studies could investigate whether periodic or aperiodic abnormalities depend on the location and nature of the underlying pathology.
Both periodic band power and aperiodic exponents influence the changes in abnormality.  While the aperiodic exponent in isolation cannot classify patient outcome in interictal data, other studies show that it is impactful for other aspects of epilepsy including seizure prediction and SOZ localization. \cite{vanHeumen2021} and \cite{Yang2023} showed that the SOZ had a steeper aperiodic offset and exponent during seizures. \cite{Liu2023} pointed out that aperiodic features especially the exponent played a decisive role in classification of stages of epilepsy, while \cite{Kundu2023} showed that aperiodic exponent decreased over time following seizure freedom in a patient who went through multiple resection. \cite{Coa2022} also found that aperiodic parameters can be applied to analyze the efficacy of vagus nerve stimulation in patients with drug-resistant epilepsy. These findings have been supported by the idea that the excitation/inhibition (E/I) (im)balance can be estimated by the use of aperiodic exponent \citep{Gao2017}. \cite{Gao2017} pinpointed that changes in the balance of excitation and inhibition (E:I balance) are associated with alterations in the aperiodic exponent within a specific frequency range (30–70 Hz).

Previous studies indicated that interictal dynamics, besides focusing on narrow-band approaches, could offer valuable insights for the identification of abnormality \citep{Chen2021,Höller2015}. Interictal spikes, sharp waves, and high-frequency oscillations (HFOs) were proposed as putative biomarkers of epileptic activity. HFOs were proposed as a superior biomarker to identify epileptogenic tissue compared to spikes \citep{Jacobs2008,Jacobs2010,vantKlooster2017}. HFOs were not as effective as spikes in guiding epilepsy surgery, overall and for temporal lobe epilepsy, and though there was a potential superiority of spikes over HFOs, HFO-guidance showed non-inferiority in the subgroup with extratemporal lobe epilepsy \citep{Zweiphenning2022}. Patients with positive outcome had a significantly higher percentage of HFO-generating areas removed compared to patients with a negative outcome \citep{Jacobs2010}. However, \cite{Roehri2018} concluded that HFOs were not superior biomarkers compared to interictal spikes and \cite{Gliske2018} found that HFO locations varied across different iEEG during sleep in most patients, spanning hours or days. While interictal spikes may manifest in the EEGs of patients, their surgical removal did not differentiate outcome groups \citep{Taylor2022}.

On the group level, neither periodic band power nor aperiodic exponent abnormalities alone distinguished patient outcome in iEEG or MEG cohorts. In both cases, we found that both periodic and aperiodic abnormalities are necessary to distinguish patient outcomes, with neither sufficient in isolation. In future studies, a quantitative multi-modal approach could be pursued to investigate abnormalities originating from periodic and aperiodic components. Possible approaches are directly comparing abnormalities across modalities or simulate multiple proposed surgeries and combining hypothetical $D_{RS}$ values from multiple modalities to determine the optimal resection, as proposed by \citep{Horsley2023}.

Our work has limitations. We only explored the specparam solution for power spectrum decomposition, while other approaches offer different mathematical solutions \citep{Gerster2022,Wilson2022}. Additionally, we limited our analysis to the power spectrum up to 30Hz to exclude potential issues with line noise and error-prone curve fitting at higher frequencies. Consequently, we excluded gamma band abnormalities from our analysis. While we used a straightforward method to measure the distinguishability between resected and spared tissue, there are other measures that solely consider the severity and size of abnormalities in resected tissue \citep{Owen2023}. Additionally, we only focused on localized abnormality while network based approaches may introduce more detailed insight into epileptogenic abnormality \citep{Bernabei2022,lagarde2018,Li2018,Shah2019,Sinha2021}. Finally, we only considered one time period in our analysis, while it is important to understand the temporal aspects of these properties. However, \cite{Wang2023} showed that $D_{RS}$ is a relatively robust metric across time, which suggests that our findings may also be temporally robust. Additionally, \cite{Wiesman2022} described the minimum amount of duration of recording necessary to obtain reliable estimates of both periodic and aperiodic components. While we used only 70 second epochs, they recommend longer time periods of MEG data (2-3 minutes, depending on the frequency band) to reliably estimate these features. Future studies are needed to investigate how different levels of signal to noise ratio influence these time intervals and if these recommendations also apply to iEEG data.

In conclusion, our study reveals that the type of spectral component providing the most information about abnormality location and patient outcome varies among patients. Consideration of both periodic and aperiodic abnormalities are necessary to distinguish spectral abnormalities. These findings further our goal to identify epileptogenic abnormalities and develop reliable methods to predict patient outcomes.

\newpage

\section{Acknowledgements}
We thank members of the Computational Neurology, Neuroscience \& Psychiatry Lab (www.cnnp-lab.com) for discussions on the analysis and manuscript; C.K. is supported by Epilepsy Research UK Foundation, P.N.T. and Y.W. are both supported by UKRI Future Leaders Fellowships (MR/T04294X/1, MR/V026569/1). JD, JdT are supported by the NIHR UCLH/UCL Biomedical Research Centre.

\newpage
\bibliography{refs}
\newpage


\renewcommand{\thefigure}{S\arabic{figure}}
\setcounter{figure}{0}
\counterwithin{figure}{section}
\counterwithin{table}{section}
\renewcommand\thesection{S\arabic{section}}
\setcounter{section}{0}

\section*{Supplementary}

\doublespacing

\section{Normative maps}

\begin{figure}[H]
	\centering
	\includegraphics{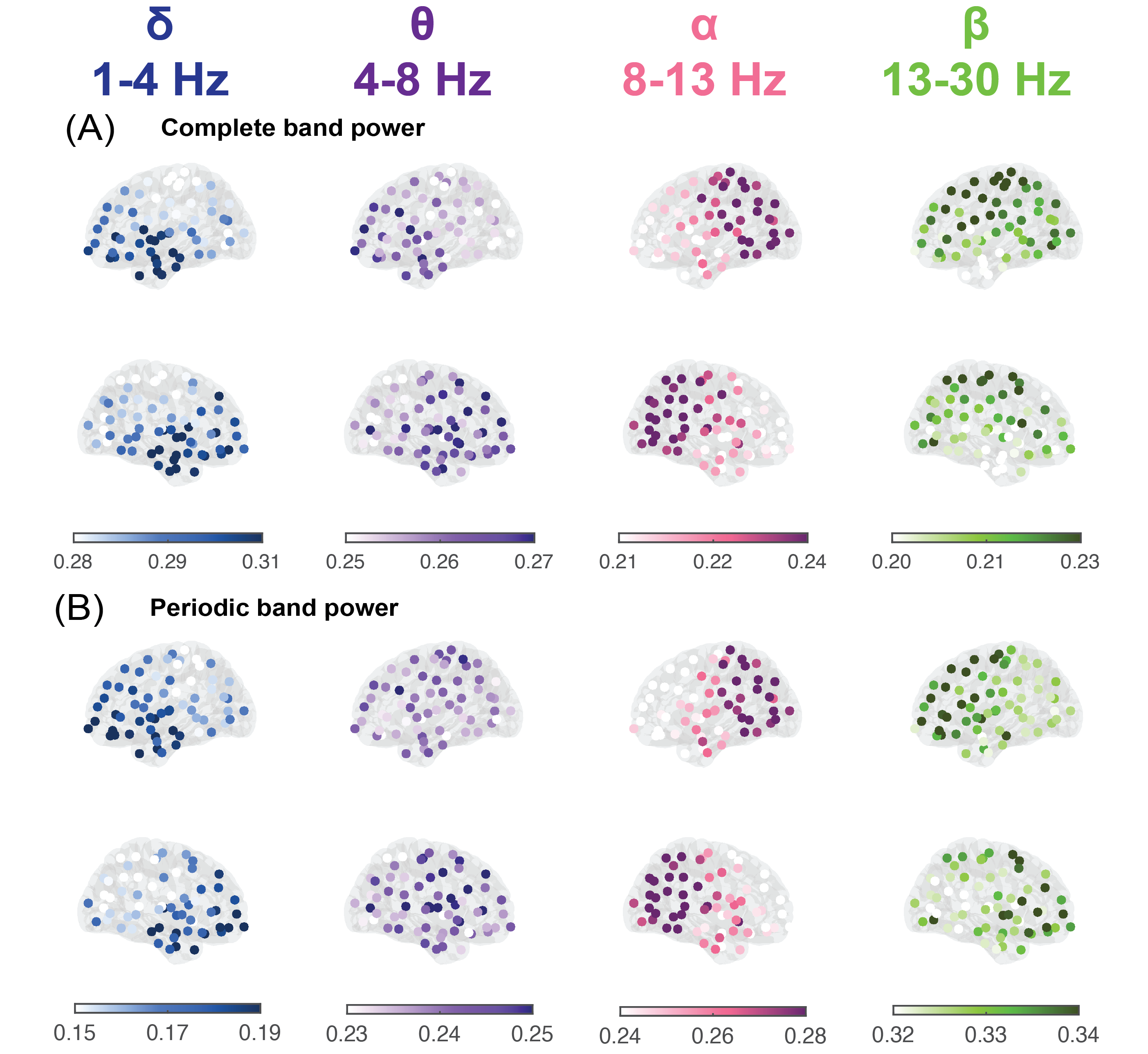}
	\caption{Side views of the normative maps generated based on (A) relative complete band power and (B) relative periodic band power}
	\label{Normative maps}
\end{figure}

Figure S1.1 shows clear patterns as described in the main results section. The anterior temporal and frontal regions have the highest delta power, while the parietal and occipital regions show significant alpha power. Periodic band power reproduces the same spatial patterns observed in complete band power.

\begin{figure}[H]
	\centering
	\includegraphics{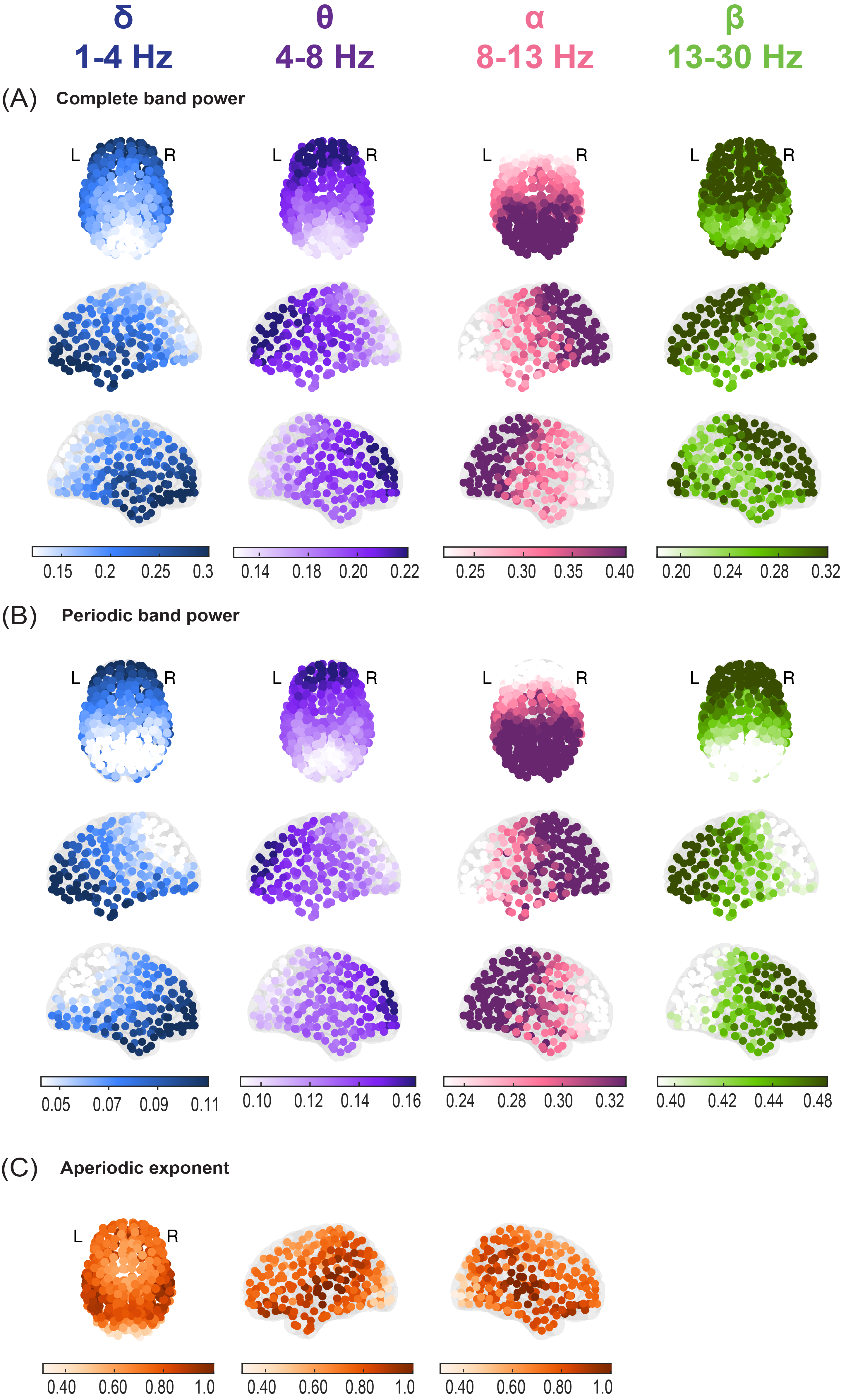}
	\caption{Normative maps of MEG relative complete band power, relative periodic band power, and aperiodic component. Spatial distribution of mean ROI values of, (A) relative complete band power, (B) relative periodic band power, and (C) aperiodic exponent}
	\label{Normative maps}
\end{figure}

Figure S1.2 depicts similar patterns as the iEEG normative maps. Namely, the anterior temporal and frontal regions have the highest delta power, while the parietal and occipital regions show significant alpha power. Periodic band power produces similar spatial patterns. In the aperiodic exponent maps, we found higher values in the posterior section of the temporal lobe and the parietal lobe. As mentioned in the main text, the MEG data set we used was eyes-closed recording, which may reduce exponent values in the occipital lobe.

\section{Correlation between abnormalities}

\begin{figure}[H]
	\centering
	\includegraphics{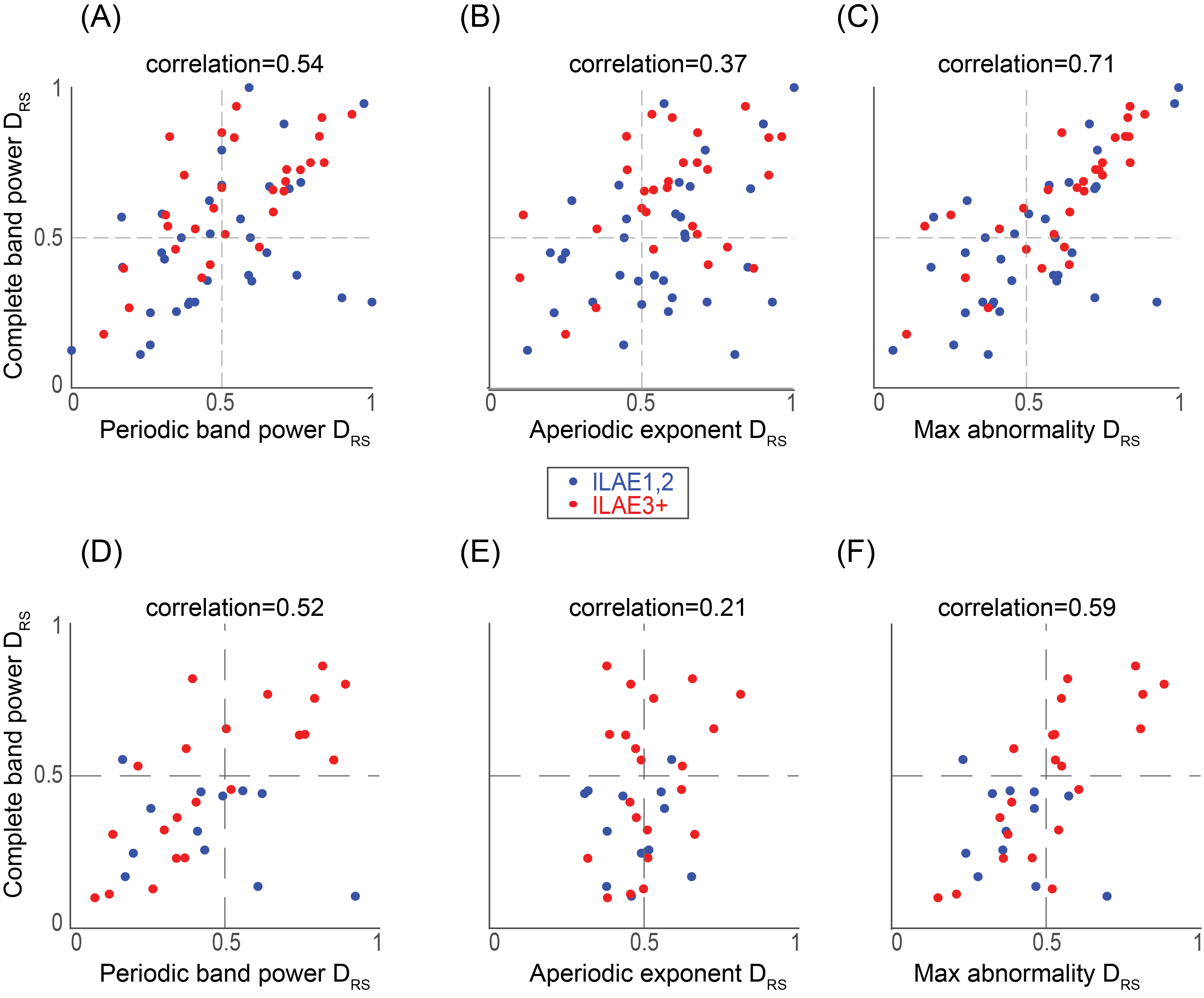}
	\caption{(A-C) Distribution of $D_{RS}$ values obtained from the iEEG complete band power plotted against $D_{RS}$ values based on abnormalities calculated from (A) periodic band power, (B) aperiodic exponent, and (C) maximum abnormalities selected from periodic band power and aperiodic exponent abnormalities for each ROI. Each data point in the plot represents an individual patient. Spearman’s $\rho$ was used to quantify the correlation between each set of $D_{RS}$ values. (D-F) Distribution of $D_{RS}$ values obtained from the MEG complete band power plotted against $D_{RS}$ values based on abnormalities calculated from (D) periodic band power, (E) aperiodic exponent, and (F) maximum abnormality based on periodic band power and aperiodic exponent.}
	\label{Normative maps}
\end{figure}

Figure S2.1 shows that selecting the higher abnormality score across periodic band power and aperiodic component solutions yield  $D_{RS}$ scores that are closely correlated with the $D_{RS}$ scores calculated from abnormalities based on the complete band power solution on the group level both in the iEEG and MEG cohorts.

\section{Proportion of features selected}

\begin{figure}[H]
	\centering
	\includegraphics{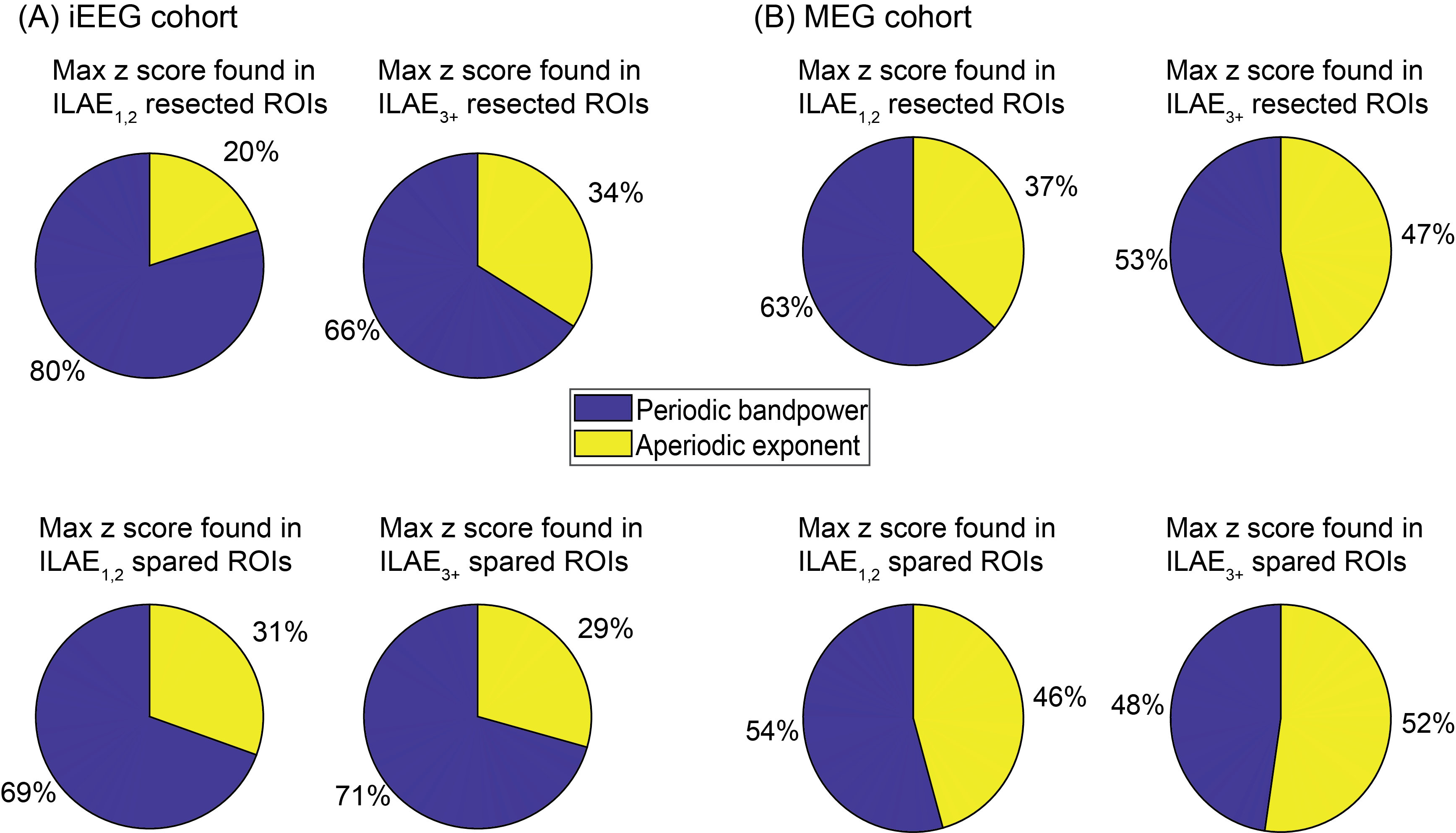}
	\caption{Proportion of maximum abnormality found from periodic band power and aperiodic exponent in (A) iEEG cohort and (B) MEG Cohort. .
	}
	\label{Normative maps}
\end{figure}

\begin{figure}[H]
	\centering
	\includegraphics{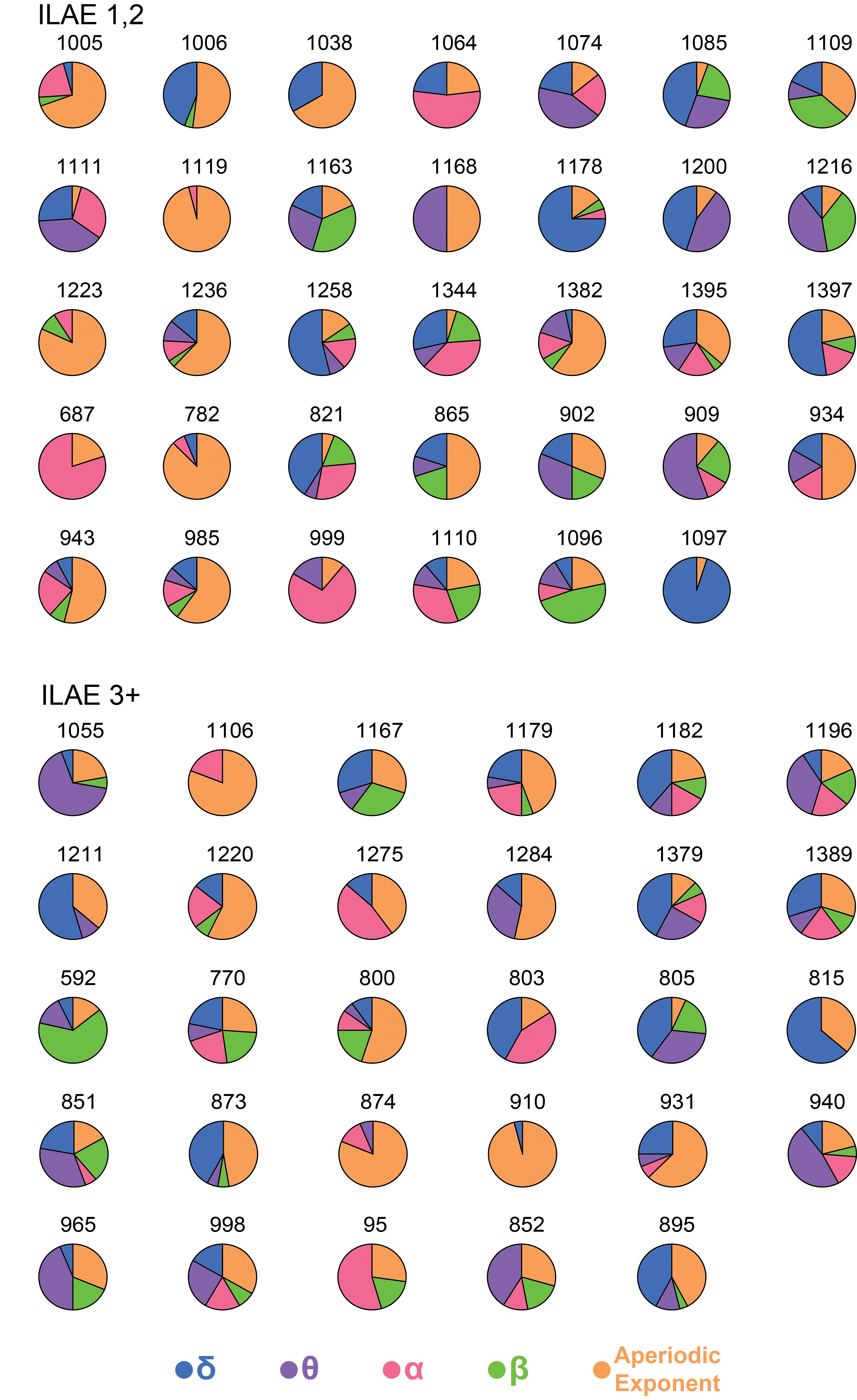}
	\caption{Proportion of maximum abnormality found from each frequency band of the periodic band power and aperiodic exponent (iEEG cohort), separated by patient outcome.
	}
	\label{Normative maps}
\end{figure}

\begin{figure}[H]
	\centering
	\includegraphics{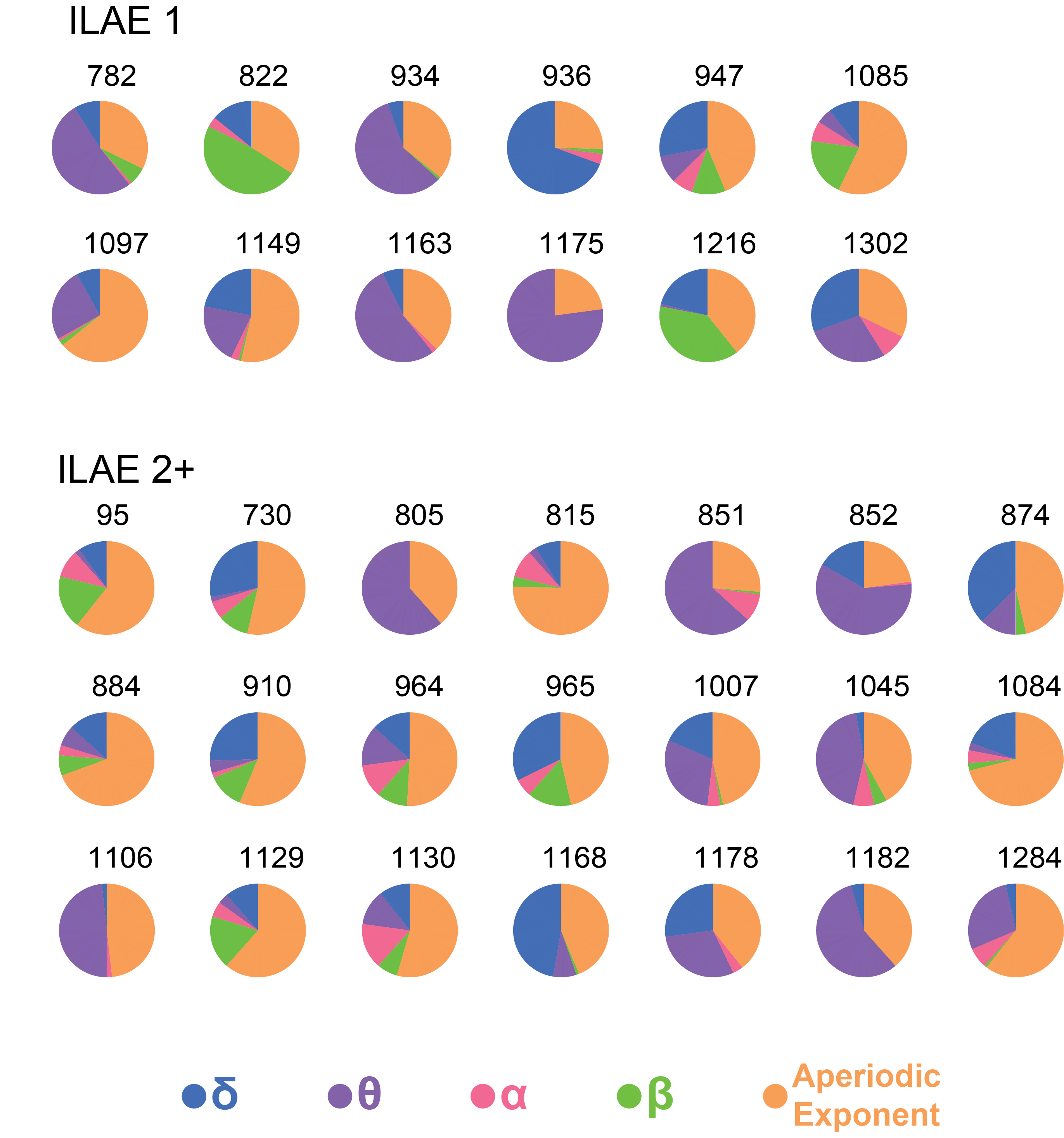}
	\caption{Proportion of maximum abnormality selected from each frequency band of the periodic band power and aperiodic exponent (MEG cohort), separated by patient outcome. 
	}
	\label{Normative maps}
\end{figure}

Figures S3.1, S3.2 and S3.3 indicate that abnormalities may be driven by different components (different frequency bands and aperiodic exponent) in different patients and in different regions. We therefore next computed the maximum abnormality, irrespective of the component used (periodic or aperiodic). 

\section{Max abnormality selected in two example patients}

 \begin{figure}[H]
	\centering
	\includegraphics[width=\textwidth]{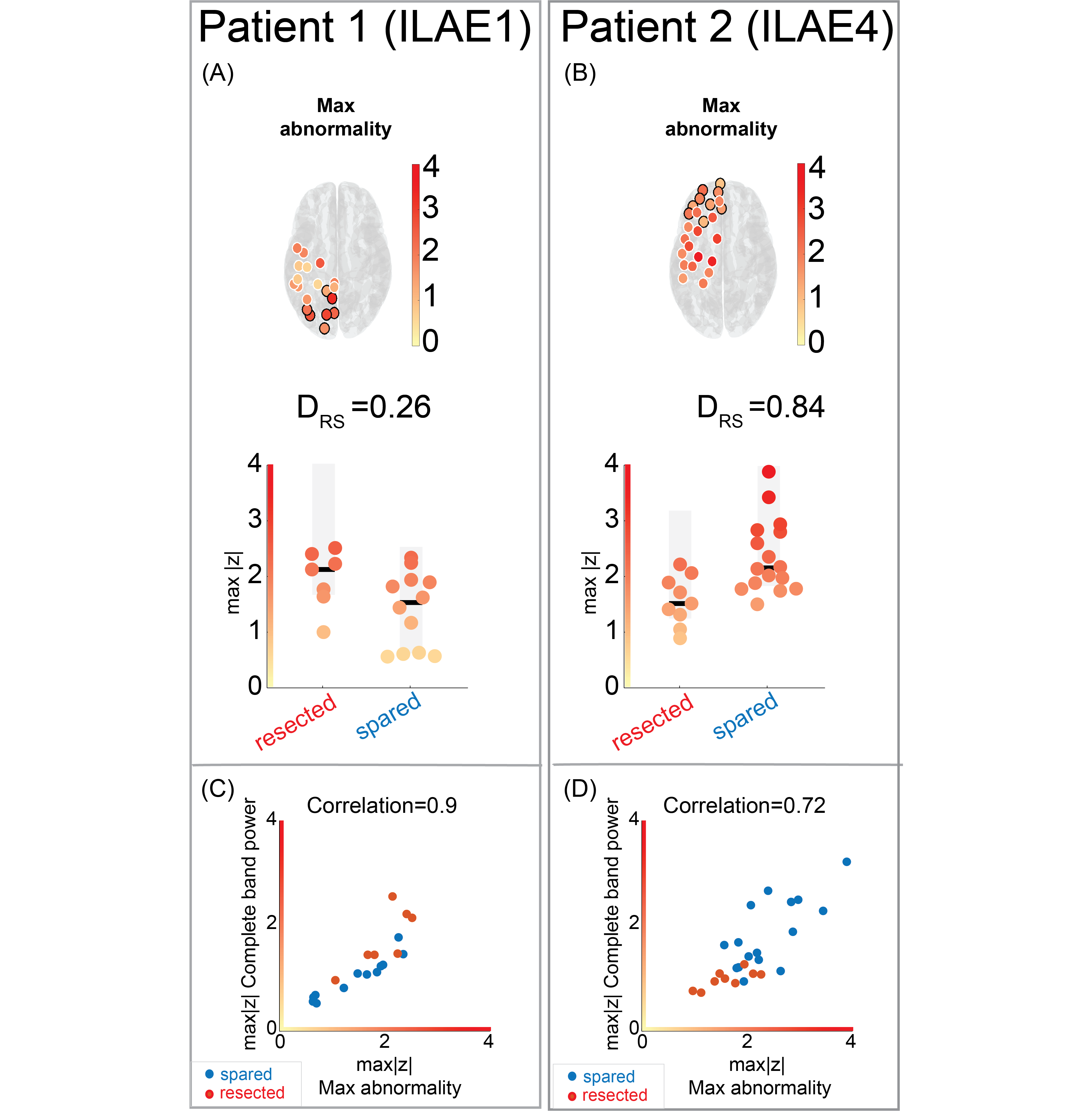}
	\caption{Max abnormality found in two example patients. Here we show the case when we select the higher abnormality score between periodic band power and aperiodic exponent for every ROI in a $ILAE_1$ (A) and and an $ILAE_4$ (B) patient. Correlation calculated between the max abnormality scores and complete band power abnormality scores. In both patients (C-D), maximum abnormalities correlate high with complete band power abnormalities.  
	}
	\label{Normative maps}
\end{figure}
Fig. S4 demonstrates the high similarity between maximum abnormality values and complete band power abnormalities across all ROIs. This similarity is also reflected in the correlation between these measures in the two example patients analyzed initially.
\end{document}